\documentclass[prb,aps,twocolumn,showpacs,superscriptaddress,amsmath,amssymb]{revtex4}
\usepackage{graphicx}
\usepackage{subfigure}
\usepackage{hyperref}
\usepackage{dcolumn}
\usepackage{multirow}
\usepackage{bm}
\usepackage{bbold}
\usepackage[capitalize]{cleveref}

\begin{document}
\title{Fractional quantization of the topological charge pumping in a one-dimensional superlattice}

\author{Pasquale Marra}
\email[]{pasquale.marra@spin.cnr.it}
\affiliation{CNR-SPIN, I-84084 Fisciano (Salerno), Italy}
\author{Roberta Citro}
\affiliation{CNR-SPIN, I-84084 Fisciano (Salerno), Italy}
\affiliation{Dipartimento di Fisica ``E. R. Caianiello'', Universit\'a di Salerno, I-84084 Fisciano (Salerno), Italy}
\author{Carmine Ortix}
\affiliation{Institute for Theoretical Solid State Physics, IFW Dresden, D-01069 Dresden, Germany}


\begin{abstract}
A one-dimensional quantum charge pump transfers a quantized charge in each pumping cycle.
This quantization is topologically robust being analogous to the quantum Hall effect.
The charge transferred in a fraction of the pumping period is instead generally unquantized.
We show, however, that with specific symmetries in parameter space the charge transferred at well-defined fractions of the pumping period is quantized as integer fractions of the Chern number.
We illustrate this in a one-dimensional Harper-Hofstadter model and show that the fractional quantization of the topological charge pumping is \emph{independent} of the specific boundary conditions taken into account.
We further discuss the relevance of this phenomenon for cold atomic gases in optical superlattices.
\end{abstract}

\maketitle

\newcommand{\nodag}{{\phantom{\dag}}}

\section{Introduction}
Three decades ago, it was established that the quantized Hall conductance of the integer quantum Hall effect (QHE)~\cite{Klitzing1980} is a topological invariant~\cite{Thouless1982} classifying the ground state of the system.
The recent discovery of topological insulators in both two-dimensional (2D)~\cite{Kane2005,Koenig2007} and three-dimensional (3D) systems~\cite{Hasan2010,Qi2011} has enormously boosted the interest in topologically non-trivial states of matter.
Apart from these presently much studied 2D and 3D states of matter, there is another quantum phenomenon of topological origin: the one-dimensional (1D) quantum charge pump.
In systems typically described by a 1D Harper model~\cite{Harper1955,Aubry1980,Han1994}, charge transfer is induced by a periodic potential which evolves adiabatically in time.
Thouless~\cite{Thouless1983} has shown that the amount of charge pumped in one cycle can be expressed in terms of the Chern invariant precisely as in the 2D QHE.
The strict analogy among these two phenomena is also reflected in the profound relations among the 2D QHE Hamiltonian and the 1D Harper model and its off-diagonal variant~\cite{Harper1955,Aubry1980,Thouless1983,Han1994,Ho2012}.

The experimental realization of a 1D Harper system has been implemented, e.g., in photonic waveguide arrays~\cite{Kraus2012a} and in optical lattices~\cite{Aidelsburger2013,Miyake2013}.
In the former, the periodic modulation of the on-site potential was produced by controlling inter-waveguide distances.
By adiabatically varying the relative phase between the modulation and the underlying lattice, light was pumped across the sample, revealing the topological nature of the pumping.
In optical lattice experiments instead, the Harper Hamiltonian has been modeled using the interaction between cold atoms and lasers.
As reported in Refs.~\onlinecite{Aidelsburger2013,Miyake2013}, rubidium atoms were loaded in an optical lattice and external lasers were used to coax the atoms into circular motion, analogous to the motion of electrons in a magnetic field.
These advances on the experimental side renewed the interest~\cite{Kraus2012a,Wang2013,Lang2012} in the theory of the quantum charge pump.

In this work we uncover an important and observable property of quantum charge pumps with additional symmetries in the parameter space: at well-defined fractions of the pumping period the system transfers a charge which is quantized in fractions of the Chern invariant of the bulk system.
This phenomenon originates from the symmetries of the gauge-invariant Berry curvature in momentum space and is not related to the emergence of fractional charges in the system.
On top of this, an explicit calculation for systems with open boundary conditions shows that the fractional quantization of the charge pumped \emph{does not} rely on the presence of translational symmetry, which can be interpreted as the prime physical consequence of the topological nature of this phenomenon.

\begin{figure}[t]
\centering
\includegraphics[scale=1,resolution=600]{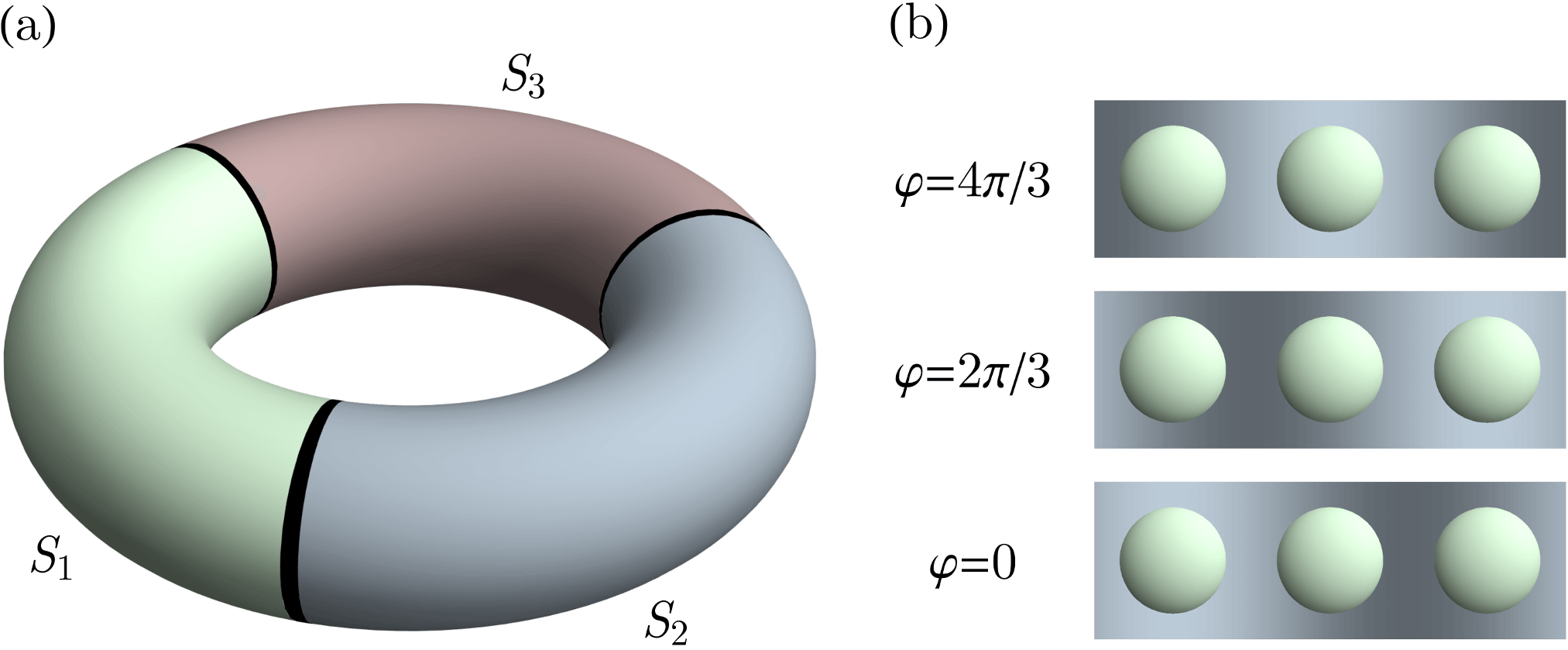}
\subfigure{\label{fig:PartitionTorus}}
\subfigure{\label{fig:System}}
\vspace{-12pt}
\caption{
(a)
Partition of the parameter space into three subsets $S_n$ which are equivalent up to a gauge transformation.
(b)
A 1D superlattice with a periodic perturbation $\propto\cos{(2\pi R_i/3+\varphi)}$.
}
\end{figure}%

\section{Topological charge pumped in periodic systems}

In a Bloch system the Chern number of an occupied band is proportional to the integral of the Berry curvature over the whole Brillouin zone~\cite{Xiao2010,Qi2011}.
The Chern number is topologically invariant and is quantized as an integer number provided there is a full gap separating the occupied bands from the unoccupied ones.
Let us then consider the case where additional point or space group symmetries induce a partitioning of the Brillouin zone into subregions which are transformed one into the other under a certain group of symmetry transformations.
More precisely, one can imagine a number $q$ of simply-connected, non-overlapping, and equally-sized subsets $S_n$ which cover the whole Brillouin zone, as shown in \cref{fig:PartitionTorus}, and such that the Bloch states $\vert\psi({\bf k})\rangle$ are equivalent up to a gauge transformation $U({\bf k})$, that is $\vert\psi({\bf k}')\rangle=U({\bf k})\vert\psi({\bf k})\rangle$ with ${\bf k}'\in S_n$ and ${\bf k}\in S_1$.
Since the Berry curvature is a gauge-invariant measurable quantity --- it is defined as $\Omega_i({\bf k})=[\nabla_{\bf k}\times {\bf A}_i({\bf k})]\cdot \hat{\bm z}$, where ${\bf A}_i({\bf k})=-\imath\langle\psi({\bf k})\vert\frac{\partial}{\partial{\bf k}}\vert\psi({\bf k})\rangle$ is the Berry connection --- the total Chern number can be evaluated as a sum of integrals of the Berry curvature over each of the subsets $S_n$, which all give an equal contribution, and therefore
\begin{equation}\label{ChernPartition}
C_j=
-\frac{q}{2\pi}\sum_{i\leq j}
\int_{S_n}{\rm d}k^2\ \Omega_i({\bf k})
.
\end{equation}
In 2D solids the integral of the Berry curvature over the individual subsets $S_n$ is not associated with any meaningful physical quantity.
This, however, is not generally the case for time-dependent systems.
Let us consider, for instance, a 1D band insulator under the action of an adiabatic time-varying perturbation, periodic in time and depending on a parameter $\varphi(t)$ with period $T$.
If the chemical potential is inside a gap for any possible value of $\varphi(t)$, the total charge transferred, i.e., the number of particles transported across any section of the chain at the time $t$, can be expressed as an integral over the Brillouin zone~\cite{Thouless1983,Bohm2003-295-299}
\begin{equation}\label{ChernPhi}
Q(t)=\frac{\imath}{\pi}\sum_{i\leq j}
\int_{\varphi(0)}^{\varphi(t)}{\rm d}\varphi \int_{\rm BZ} {\rm d}k
\ {\rm Im}
\left\langle \frac{\partial \psi_i }{\partial \varphi} \Bigl\vert \frac{\partial \psi_i }{\partial k} \right\rangle
,
\end{equation}
where $j$ is the total number of filled bands of the system and $\vert{\psi_i}\rangle=\psi_i(k,\varphi)c^\dag_i\vert{0}\rangle$ the $i$th band eigenstate, being $c^\dag_i$ the creation operators acting on the ground state $\vert{0}\rangle$.
The integral in \cref{ChernPhi} can be related to the Zak phase~\cite{Zak1989} of the system while the charge transferred can be written in terms of the variation of the charge polarization of the 1D system~\cite{Vanderbilt1993,Xiao2010,Resta2007} along the path $\Delta\varphi=\varphi(t)-\varphi(0)$ (cf. \cref{app:polarization}).

\begin{figure}[t]
\centering
\includegraphics[scale=1,resolution=600]{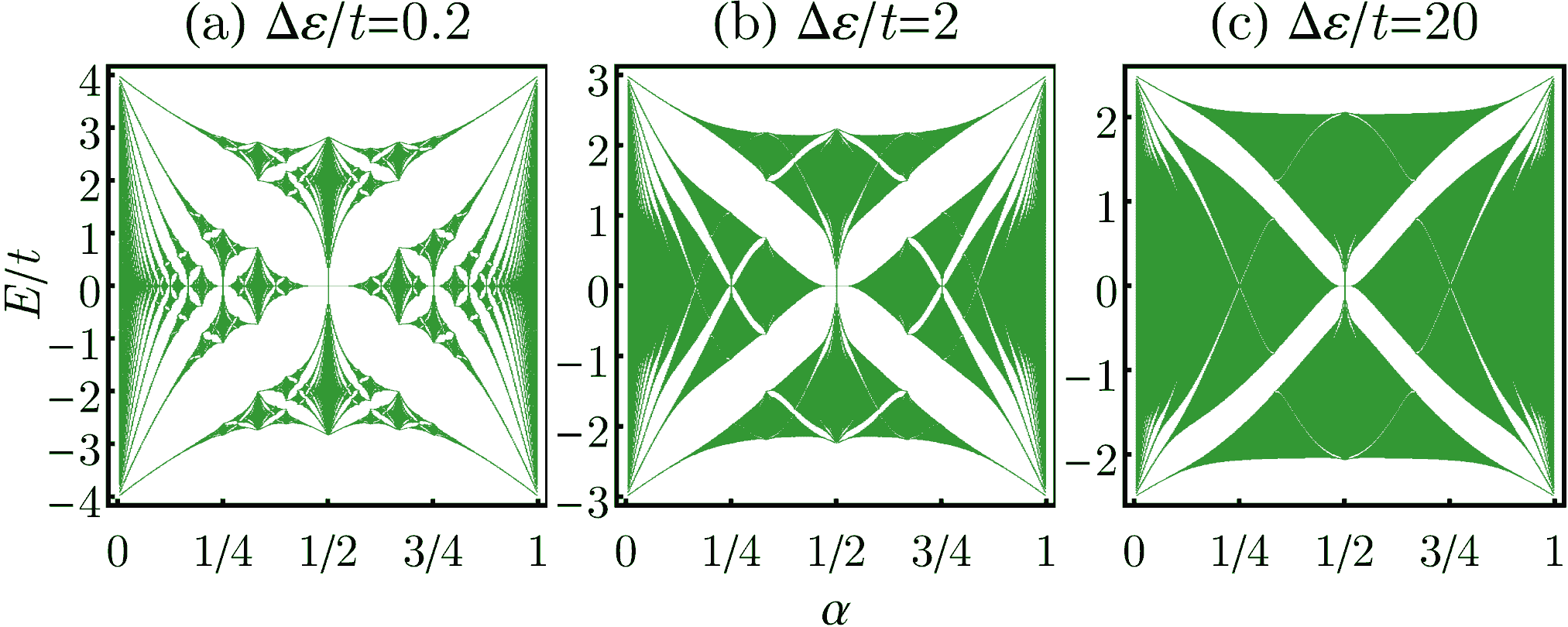}
\subfigure{\label{fig:Spectrum1}}
\subfigure{\label{fig:Spectrum05}}
\subfigure{\label{fig:Spectrum025}}
\vspace{-12pt}
\caption{
Spectra of the 1D superlattice with the modulation phase $\varphi$ varying in the interval $[0,2\pi]$, for rational values $\alpha=p/q$ and for different values of the substrate on-site energy modulation $\Delta\varepsilon$.
}
\label{fig:Spectra}
\end{figure}%

Since the underlying time-dependent Hamiltonian of the system is periodic both in the parameter $\varphi$ and in the 1D momentum $k$, it follows that, for an evolution over a full period $T$, the parameter space corresponds to a torus and $2\pi Q$ is nothing but the Berry phase over this closed manifold.
However, the integral over the full torus can be seen as being composed of $q$ non-overlapping subsets, as shown in \cref{fig:PartitionTorus}, in which the momentum $k$ traces a circle while the parameter $\varphi \in \left[\varphi(0) , \varphi(T/q) \right]$.
If the symmetries of the Hamiltonian in the parameter space mandate that all these integrals give an equal contribution, then it follows that the charge transferred in the $T/q$ fraction of the pumping period is $Q (T/q)=C_j / q$.
Hence the partitioning of the torus in the parameter space combined with the symmetry of the Berry curvature leads to a well-defined physical response: a fractionally quantized pumped charge.
This fractional quantization is robust against any perturbation preserving the symmetries of the parameter space.

Next we show how such a situation can be achieved in a time-varying 1D superlattice generated by an external potential periodic in real space.
Specifically, we consider a system of spinless particles at zero temperature confined in a 1D lattice, like the one in \cref{fig:System}, subject to a weak external potential with period $1/\alpha$ times the lattice parameter.
The system is described by the tight-binding Hamiltonian
\begin{equation}
{\cal H}=\sum_i {\Delta\varepsilon}\cos{(2\pi\alpha R_i+\varphi)} c_i^\dag c_i^\nodag - \sum_{<i,j>} t\, c_i^\dag c_j^\nodag ,
\label{TBSpace}
\end{equation}
where $\Delta\varepsilon$ is the characteristic perturbation strength while $\varphi$ is a phase which we assume to be linear in time.
We note here that this Hamiltonian describes either spinless fermions~\cite{Wang2013} or hardcore bosons in 1D optical superlattices.
In momentum space, the tight-binding Hamiltonian~(\ref{TBSpace}) becomes
\begin{equation}
{\cal H}=
\sum_{k}
-2t\cos{k}\ c_{k}^\dag c_{k}^\nodag
+\frac{\Delta\varepsilon}{2}e^{\imath\varphi}c_{{k}+{2\pi\alpha}}^\dag c_{k}^\nodag+\text{h.c.}
,
\label{TBMomentum}
\end{equation}
which is not diagonal with respect to Bloch states, since the substrate potential couples states with momenta ${k'-k}={2\pi\alpha}$.

The non-trivial topological properties of the Hamiltonian~(\ref{TBMomentum}) can be revealed by considering the external perturbation to be commensurate with the lattice, i.e., $\alpha=p/q$ with $p<q$ coprimes, leading to a superlattice with $q$ atoms in the unit cell and a corresponding mini-Brillouin zone $k \in [0,2\pi/q]$.
\Cref{fig:Spectra} shows the ensuing energy spectra obtained by diagonalizing the Hamiltonian~(\ref{TBMomentum}) with the phase $\varphi$ varying in the interval $[0,2\pi]$, for rational values of $\alpha$ and different values of the substrate modulation intensity $\Delta\varepsilon$.
The energy spectrum in the case $\Delta\varepsilon/t=2$ [\cref{fig:Spectrum1}] coincides with the well known Hofstadter spectrum~\cite{Hofstadter1976} of a 2D electron system on a square lattice subject to a uniform magnetic field oriented perpendicularly to the lattice plane.
Similarly, the energy spectra obtained for $\Delta\varepsilon/t\neq2$ [\cref{fig:Spectrum05,fig:Spectrum025}] correspond to the Hofstadter spectra in the case of rectangular lattices.
This result is an immediate consequence of the mapping~\cite{Lang2012} between the lattice version of the integer QHE problem and the 1D Hamiltonian~(\ref{TBMomentum}).
Since any of the $q-1$ gaps of the Hofstadter butterfly corresponds to a gapped state of the 1D system in the full torus spanned by $k$ and $\varphi$, one can identify the charge transferred of the 1D system with the Chern number $C_j$ labeling the $j$th gap of the Hofstadter butterfly~\cite{Thouless1982} and given by the unique integer solution $|C_j|<q/2$ of the Diophantine equation~\cite{Osadchy2001} $mq-p\,C_j=j$.

\begin{figure}[t]
\centering
\includegraphics[scale=1,resolution=600]{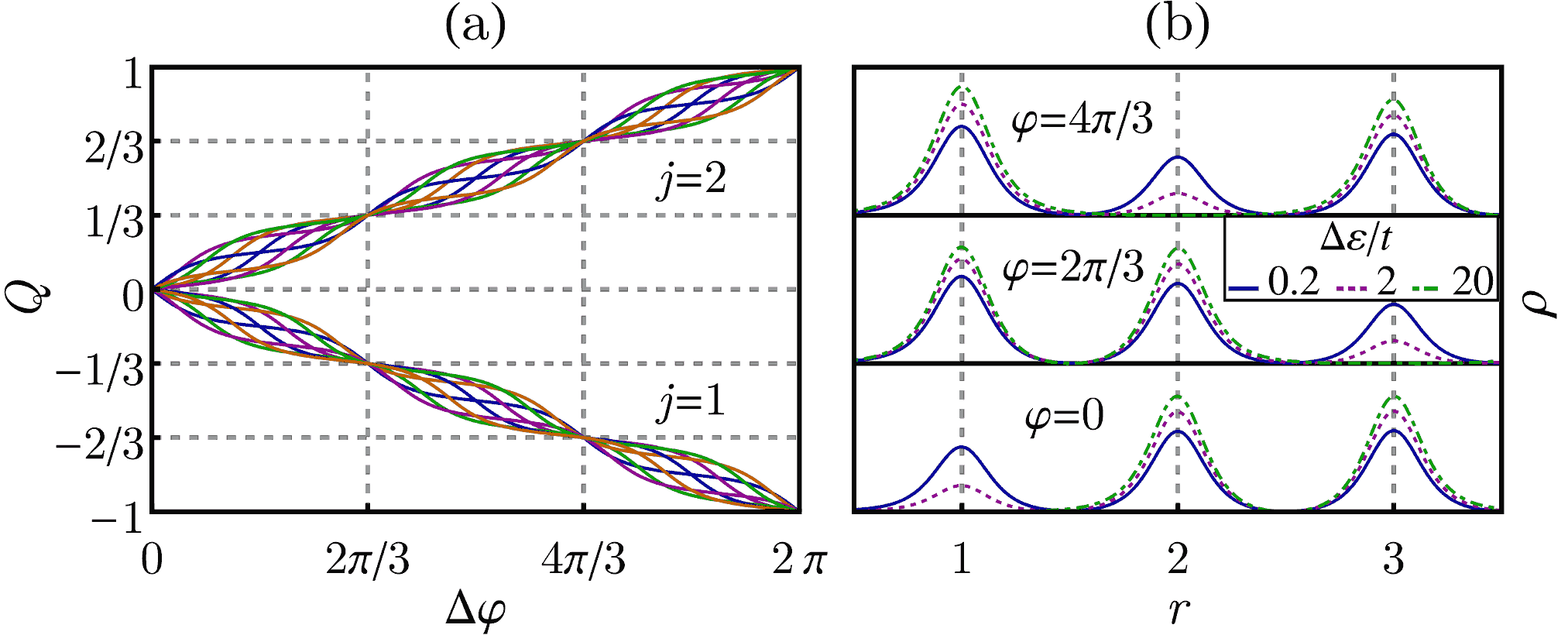}
\subfigure{\label{fig:Charge}}
\subfigure{\label{fig:Density}}
\vspace{-12pt}
\caption{
(a) Charge transferred $Q$ as a function of the modulation phase variation $\Delta\varphi$ for $p/q=1/3$, with $\Delta\varepsilon/t=2$, and for the two intraband gaps $j=1$, and 2 with Chern numbers $C_j=\mp1$ respectively, calculated via  numerical integration of \cref{ChernPhi}.
Different curves correspond to different initial values of the modulation phase $\varphi_0$.
The charge transferred along the paths $\Delta\varphi=2\pi n/q$ is quantized as fractions of the Chern number $Q=n/q C_j$, according to \cref{fractionalcharge}.
(b) Charge density $\rho$ (arbitrary units) as a function of the position $r$ over three lattice sites for different values of the phase $\varphi=2\pi n/q$ for the intraband gap $j=1$.
Different curves correspond to different regimes, in particular the kinetic ($\Delta\varepsilon/t\ll 1$, delocalized particles) and the atomic regime ($\Delta\varepsilon/t\gg 1$, particles localized on lattice sites).
}
\end{figure}

Having established the quantized particle transport in the time-varying 1D systems, we now proceed to reveal the symmetry of the Hamiltonian~(\ref{TBSpace}) which leads to a fractional quantization of the pumped charge.
To do so, we introduce the translation operator $T(n)$ which translates the whole system by an integer number of lattice sites.
It is possible to verify, by a direct substitution in the Hamiltonian~(\ref{TBSpace}), that the translation of $n$ lattice sites is equivalent to a change in the modulation phase of $\Delta\varphi=2\pi\alpha n$, i.e., $T(n){\cal H}(\varphi)T(-n)={\cal H}(\varphi+2\pi\alpha n)$.
Let us now consider a variation of the modulation phase $\Delta\varphi=2\pi n/q$ with $0<n<q$ integer.
From the Diophantine equation $mq-p\,C_n=n$ giving the Chern number $C_n$ of the $n$th gap, one obtains $2\pi n/q=2\pi m-2\pi\alpha C_n$ and therefore a change in the modulation phase $\Delta\varphi=2\pi n/q$ is equivalent to a translation of $-C_n$ lattice sites
\begin{equation}
\label{eq:unitary}
{\cal H}(\varphi+2\pi n/q)=T(-C_n){\cal H}(\varphi)T(C_n).
\end{equation}
Hence, the eigenstates of the Hamiltonian are periodic up to an arbitrary phase, i.e., $\vert\psi_i(r,\varphi+2\pi n/q)\rangle\propto\vert\psi_i(r-C_n,\varphi)\rangle$ for any energy level.
The unitary transformation in \cref{eq:unitary} corresponds to a magnetic translation of the guiding center of the Landau levels~\cite{Bernevig2013} in the original 2D Hofstadter model.
Therefore any variation in the phase modulation of a number $n$ of periods $2\pi/q$ is equivalent to a translation of exactly $-C_n$ lattice sites of all the eigenstates of the system, independent of the band index $i$.
This is immediately manifested in \cref{fig:Density} where the charge densities $\rho(r,\varphi)=\sum_{i\leq j}\vert\psi_i(r,\varphi)\vert^2$ arising from different configurations of the system $\varphi=2\pi n/3$ [cf. \cref{fig:System}] are all equivalent up to lattice translations.
Consequently, the gauge-invariant Berry curvature is periodic in the parameter space with period $2\pi/q$ and thus any adiabatic evolution $\varphi\rightarrow\varphi+2\pi/q$ contributes equally to the integral in \cref{ChernPhi}, for any initial modulation phase $\varphi$.
The total transferred charge over the cycle $2\pi$ is thus given by $Q(2\pi)=q Q(2\pi/q)$.
Put differently, the charge transferred in any of the adiabatically gapped phases, described by an adiabatic evolution with $\Delta\varphi=2\pi n/q$, is quantized in multiples of a fraction $n/q$ of the Chern number, i.e.,
\begin{equation}\label{fractionalcharge}
Q\left(\frac{2\pi n}q\right)=\frac{n}q C_j .
\end{equation}
This result is confirmed by a direct calculation of \cref{ChernPhi}, where the eigenstates have been obtained via exact diagonalization of the Hamiltonian~(\ref{TBMomentum}).
\Cref{fig:Charge} shows the calculated charge transferred as a function of the modulation phase variation $\Delta\varphi$ for $p/q=1/3$.
While the charge transferred in any of the the adiabatically gapped phase $j$ along the path $\Delta\varphi=2\pi$ is quantized as $Q=C_j$, the charge transferred along $\Delta\varphi=2\pi/q$ is quantized as a fraction of the Chern number according to \cref{fractionalcharge}, \emph{independent} of the initial phase.
As we show in the \cref{app:robust}, this result is robust against perturbations preserving the symmetries of the parameter space.

\section{Center of charge in finite systems}

\begin{figure}[t]
\centering
\includegraphics[scale=1,resolution=600]{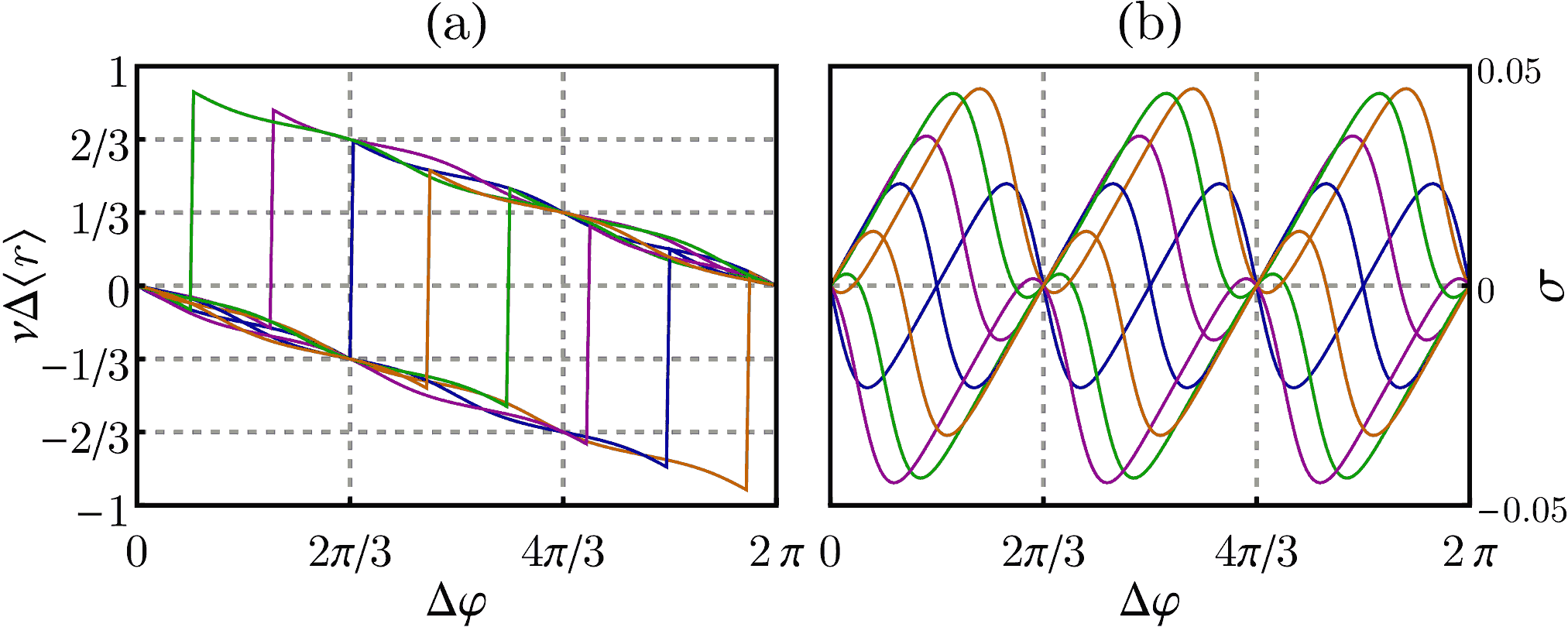}
\subfigure{\label{fig:Center}}
\subfigure{\label{fig:sigma}}
\vspace{-12pt}
\caption{
(a) Variation of the center of charge $\Delta\langle r \rangle$ as a function of the modulation phase variation $\Delta\varphi$ for a finite 1D superlattice in the thermodynamic limit, with $p/q=1/3$ and at filling $\nu=1/3$.
(b) Edge charge $\sigma$ in the thermodynamic limit, as defined in \cref{eq:CCQ}, as a function of the modulation phase variation $\Delta\varphi$.
The edge charge vanishes for any $\Delta\varphi=2\pi n/q$.
Different curves correspond to different initial phase values $\varphi_0$.
}
\label{fig:Centersigma}
\end{figure}

In an infinite periodic system, the fractional quantization of the pumped charge originates simply from the fact that for any phase shift $\Delta\varphi=2\pi n /q$, the external potential configuration is shifted by an integer number $-C_n$ of lattice sites and so does the local density~\cite{note_wannier} $\rho(r,\varphi)$ [cf. \cref{fig:Density}] in the kinetic ($\Delta\varepsilon/t\ll 1$) as well as in the atomic regime ($\Delta\varepsilon/t\gg 1$)~\cite{Niu1990}.
This, however, makes no assertion on whether or how signatures of the fractional quantization of the pumped charge can be directly detected in systems with open boundary conditions, where the translational symmetry is explicitly broken.
To answer this question, we thus consider the temporal variation of the center of charge $\langle r(\varphi) \rangle=\frac{1}{N}\int_{0}^{L} {\rm d} r \rho(r,\varphi) r $ in a finite system of length $L$, which has been proposed to encode the topological pumped charge in atomic clouds~\cite{Wang2013}.
Let us consider the adiabatic evolution $\varphi(0)\rightarrow\varphi(t)$ taking place in a finite time interval.
If one multiplies $r$ to both terms of the continuity equation $\partial \rho / \partial t = \nabla J$ with $J$ the current density~\cite{Wang2013}, integrates over space and time, and assumes that the phase parameter is linear in time, one obtains 
\begin{align*}
N
&\int_{\varphi(0)}^{\varphi(t)} {\rm d} \varphi
\frac{\partial \langle r(\varphi) \rangle}{\partial \varphi} 
=\\&
\int_{\varphi(0)}^{\varphi(t)} {\rm d}\varphi \int_0^L {\rm d}r J(r,\varphi)
-L \int_{\varphi(0)}^{\varphi(t)} {\rm d}\varphi J(L,\varphi)
.
\end{align*}
In the equation above, the second and third integral correspond to the pumped charge respectively within the bulk and at the edges of the system.
As a consequence, along any adiabatic evolution $\varphi(0)\rightarrow\varphi(t)$, the variation of the center of charge $\Delta\langle r \rangle=\langle r[{\varphi(t)}] \rangle-\langle r[{\varphi(0)}] \rangle$ is proportional to the sum of the pumped charge density and of the edge charge density as
\begin{equation}
\nu
\Delta \langle r \rangle
= Q - \sigma
,
\label{eq:CCQ}
\end{equation}
where $\nu$ is the number of particles per lattice site.
\Cref{fig:Centersigma} shows the variation of the center of charge $\nu \Delta \langle r \rangle$ and the corresponding edge charge $\sigma$ calculated in the thermodynamic limit $L\rightarrow \infty$ (see \cref{app:thermo} for details) for a 1D superlattice with open boundaries and with $p/q=1/3$ and at filling $\nu=1/3$.
If the current density were to die out at infinity, the contribution of the edge charge would vanish and thus the drift in the center of charge would correspond \emph{precisely} to the value encountered in a periodic system modulo an integer number of charges.
However, as long as $t\neq 0$ and $\Delta\varepsilon\neq 0$, we find the edge charges to give a sizable contribution to the drift in the center or charge [cf. \cref{fig:sigma}] even in the thermodynamic limit $L\rightarrow \infty$.
Exceptions are encountered for phase variations $\Delta\varphi=2\pi n/q$, explicitly proving that the variation of the center of charge for $\Delta\varphi=2\pi n/q$ of the system in any of the gapped phase is quantized as fractions of the Chern number as
\begin{equation}
\nu
\Delta \langle r \rangle \left(\frac{2\pi n}{q}\right)
= \frac{n}q C_j \mod 1,
\label{fractionalcharge2}
\end{equation}
which, together with \cref{fractionalcharge}, is the main result of this work.
This result is confirmed by direct numerical calculations also in the case $q>3$ (see \cref{app:thermo}).
Therefore, the quantization of the charge pumped in an infinite system as fractions of the Chern number corresponds to the quantization of the variation of the center of charge in a finite system.
This quantization has a topological origin and cannot be related to any translational symmetry, which is in fact broken by the open boundary conditions.
It is of paramount importance to notice that the equivalence between the variation of the center of charge and the charge pumped for well-defined fractions of the phase variation $\Delta\varphi$ is not due to the complete absence of edge effects.

The experimental observation of a fractional charge pump can be achieved in the recently realized 1D optical superlattices~\cite{Aidelsburger2013,Miyake2013} confined by a box-shaped trap~\cite{Gaunt2013}.
In these systems the modulation phase $\varphi$ can be varied adiabatically in time (see, e.g., Ref.~\onlinecite{Wang2013}).
The theoretical prediction of \cref{fractionalcharge2} can be therefore validated by a measure of the variation of the center of charge along a well-defined fraction of the pumping cycle $2\pi/q$.
We emphasize that the center of charge of the superlattice can be inferred from the local density distribution\cite{Wang2013}, which is directly accessible by \emph{in situ} measurement, or from time-of-flight imaging~\cite{Bloch2012}. 
We also point out that finite temperature effects do not interfere with the quantum pumping process for temperatures smaller than the energy gap.
For a shallow optical lattice the quantum to thermal transition is expected to occur at temperatures much larger than the recoil energy $E_R$, and temperatures of the order of $0.1 E_R/k_B$ (tens of $\rm nK$) can be easily achieved in current experiments with, e.g., $^{40}\rm{K}$ atoms.

\section{Conclusions}
We have shown, in conclusion, that the charge transferred and the variation of the center of charge in 1D superlattices are quantized as fractions of the Chern number at well-defined fractions of the pumping period.
This fractional quantization has a topological nature and can be probed in current experiments with cold atoms in optical superlattices.
Demonstration of a fractional quantization of charge represents an important step towards the comprehension of topological states of matter.

\begin{acknowledgments}
P.M. would like to thank Alessandro Braggio, Jeroen van den Brink, Luca Parisi, Nicolas Regnault, and Francesco Romeo for fruitful discussions.
P.M. and R.C. acknowledge the project FIRB-2012-HybridNanoDev (Grant No. RBFR1236VV).
C.O. acknowledges the financial support of the Future and Emerging Technologies (FET) programme within the Seventh Framework Programme for Research of the European Commission, under FET-Open Grant No. 618083 (CNTQC). 
\end{acknowledgments}

\appendix
\section{Polarization in an infinite 1D system}
\label{app:polarization}

In this appendix, we show that in an infinite 1D system the variation of the polarization equates the pumped charge during an adiabatic evolution. 
Following Refs.~\onlinecite{Vanderbilt1993,Resta2007} one can define the polarization in an infinite system in terms of the Zak phase~\cite{Zak1989} as
\begin{equation}
\label{eq:P}
P(\varphi)=
-\frac{1}{2\pi}\sum_{i\leq j}\int_{\rm BZ}
{\rm d}k\ A_{i,k}(k,\varphi)
,
\end{equation}
where $A_{i,k}(k,\varphi)=-\imath\langle \psi(k)\vert\frac\partial{\partial k}\vert \psi(k)\rangle$ is the component $k$ of the Berry connection of the $i$th band. 
The variation of the polarization along an adiabatic evolution $\varphi(0)\rightarrow\varphi(t)$ is given by
\begin{equation*}
\Delta P=P[\varphi(t)]-P[\varphi(0)]=
\int_{\varphi(0)}^{\varphi(t)}{\rm d}\varphi\frac{\partial P(\varphi)}{\partial \varphi}
,
\end{equation*}
which can be written using \cref{eq:P} in terms of the Berry curvature $\Omega_i(k,\varphi)$ as~\cite{Vanderbilt1993,Xiao2010}
\begin{equation}
\label{eq:DeltaP}
\Delta P=
-\frac{1}{2\pi}\sum_{i\leq j}\int_{\varphi(0)}^{\varphi(t)}{\rm d}\varphi\int_{\rm BZ}
{\rm d}k\ \Omega_i(k,\varphi)
,
\end{equation}
with $\Omega_i(k,\varphi)=\partial_{k} A_{\varphi,i}(k,\varphi)-\partial_{\varphi} A_{k,i}(k,\varphi)$, where ${A}_{\varphi,i}(k,\varphi)=-\imath\langle\psi(k,\varphi)\vert\frac{\partial}{\partial\varphi}\vert\psi(k,\varphi)\rangle$ and ${A}_{k,i}(k,\varphi)=-\imath\langle\psi(k,\varphi)\vert\frac{\partial}{\partial k}\vert\psi(k,\varphi)\rangle$ are the two components of the Berry connection. 
We notice here~\cite{Vanderbilt1993} that both the polarization and its variations in \cref{eq:P,eq:DeltaP} are defined up to multiples of the elementary charge. 
By a comparison of \cref{eq:DeltaP} with \cref{ChernPhi} one can conclude that
\begin{equation}
\Delta P\equiv Q \mod 1
,
\end{equation}
in units of the elementary charge.
Therefore, along any adiabatic evolution $\varphi(0)\rightarrow\varphi(t)$, the variation of the polarization equates the charge pumped. 

\section{Robustness of the fractionally quantized charge pumping}
\label{app:robust}

The quantization of the charge pumped as fractions of the Chern number is robust against perturbations that preserve the symmetries of the parameter space. 
This is because the quantization mechanism relies merely on the partitioning of the parameter space in subsets which are equivalent up to a gauge transformation. 
We consider in this appendix an example of such a perturbation, and discuss briefly how this perturbation preserves the fractional quantization of the pumped charge and the variation of the center of charge at fractions of the pumping cycle. 
We consider a perturbed Hamiltonian
\begin{equation}
\tilde{{\cal H}}={\cal H}
 - \sum_{\ll i,j\gg} t'\, c_i^\dag c_j^\nodag 
,
\end{equation}
where $\cal H$ is defined as in \cref{TBSpace}, and where the perturbation is in the form of a next-nearest neighbor interaction with strength $t'$. 
This perturbation preserves the translational symmetry and the periodicity of the parameter space of the system with respect to the modulation phase $\varphi$. 
If one takes $p/q=1/3$, $\Delta\varepsilon/t=2$ and $t'\ll t$ the spectrum shows two gaps as in the unperturbed case, while at $t'\simeq 0.5 t$ the lowest energy gap closes and therefore the system is no longer topologically equivalent to the unperturbed one. 

\begin{figure}[t!]
\centering
\includegraphics[scale=1,resolution=600]{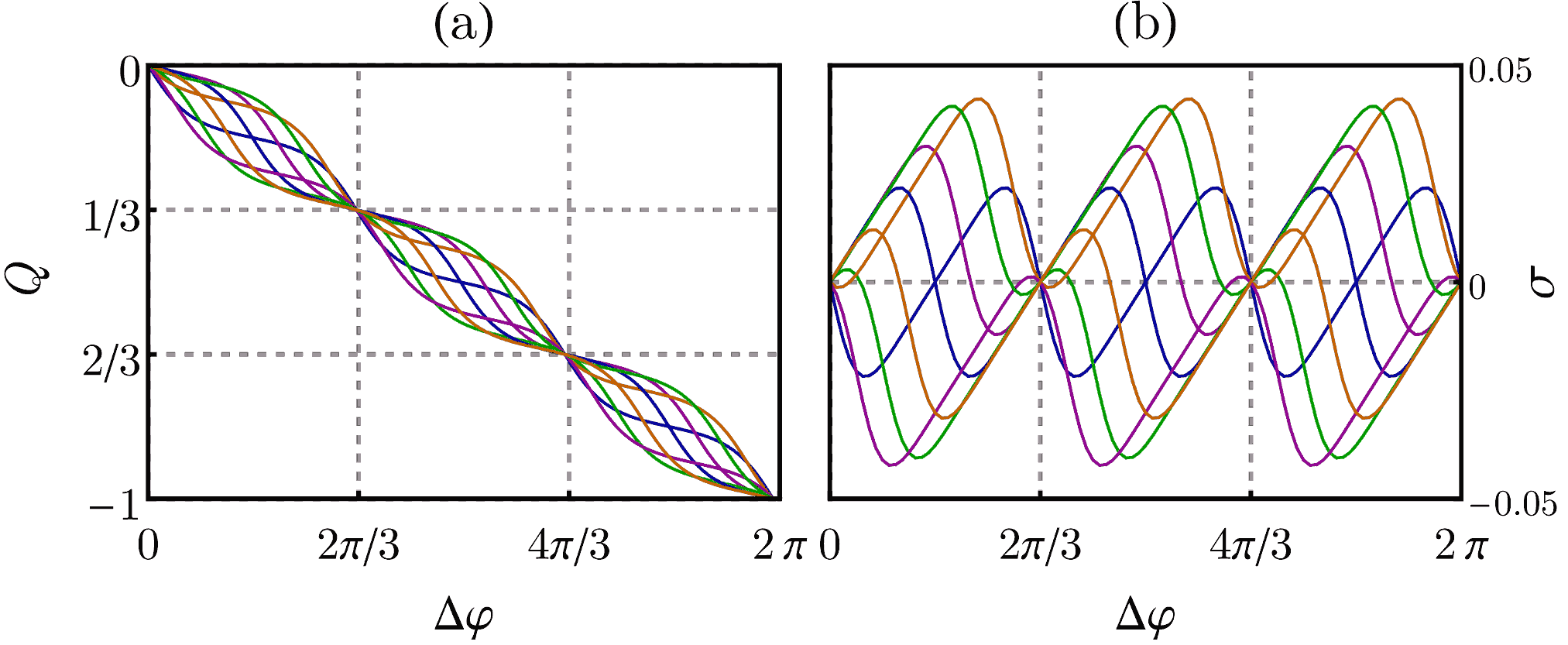}
\caption{
Charge transferred $Q$ (a) and edge charge $\sigma$ (b) in the thermodynamic limit as a function of the modulation phase variation $\Delta\varphi$ with $p/q=1/3$, at filling $\nu=1/3$, and with next-nearest neighbor perturbation $t'=0.4 t$. 
Different curves correspond to different initial phase values $\varphi$. 
}
\label{fig:Robust}
\end{figure}

\Cref{fig:Robust} shows the pumped charge and the edge charge in a finite 1D superlattice in the thermodynamic limit (see \cref{app:thermo} for details) with next-nearest neighbor interaction $t'=0.4 t$.
By a comparison with \cref{fig:Charge,fig:sigma}, one can see that the fractional quantization of the pumped charge and of the variation of the center of charge at $\Delta\varphi=2\pi n/3$ as fractions of the Chern number is preserved in the case of small perturbations.

Therefore the quantization is robust against a small next-nearest neighbor perturbation, which preserves the symmetries of the parameter space and the topology of the energy spectrum. 

\section{Center of charge in the thermodynamic limit}
\label{app:thermo}

\begin{figure}[!t]
\centering
\includegraphics[scale=1,resolution=600]{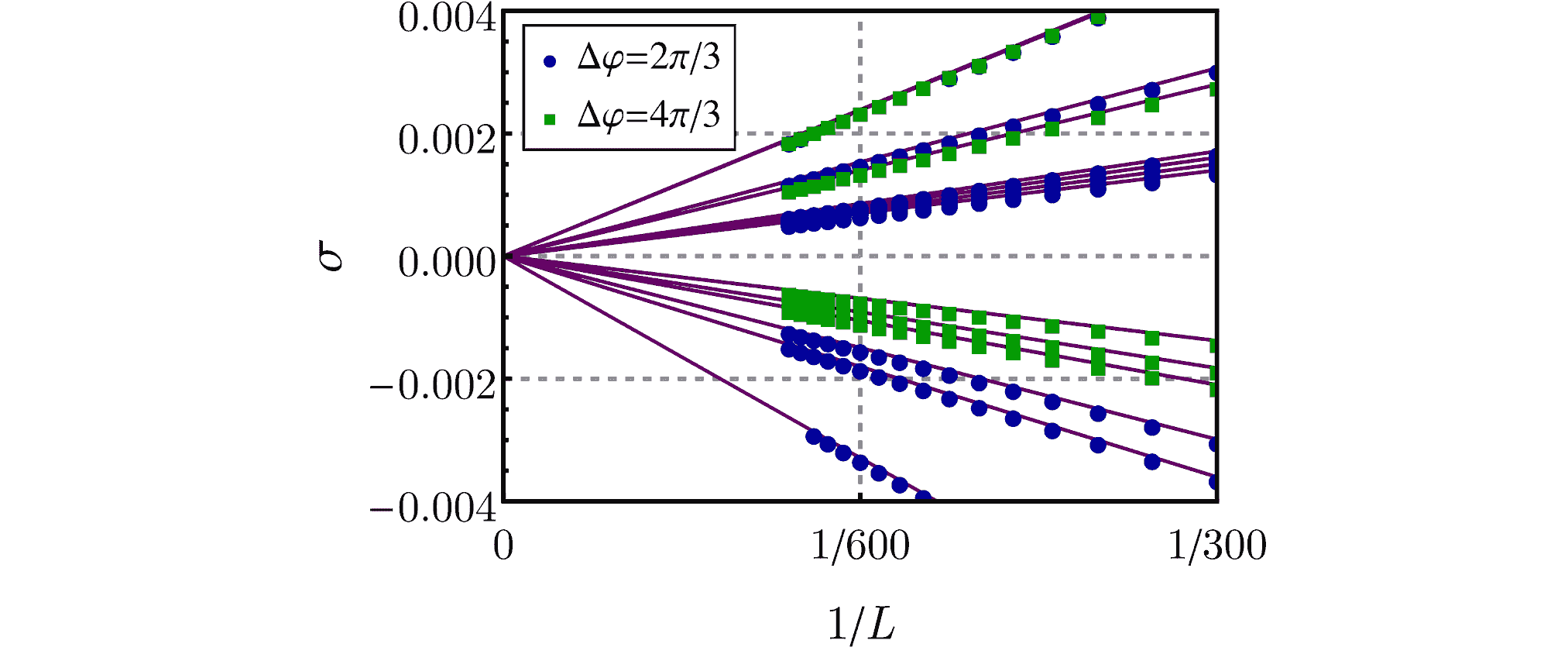}\\[-3mm]
\caption{
The calculated edge charge $\sigma$ obtained via direct calculation in a system with $p/q=1/3$ and $\Delta\varepsilon/t=2$ as a function of the system size $L$, for $\Delta\varphi=2\pi/3$ and $4\pi/3$, and for different choices of the initial phase $\varphi_0$.
The straight lines are the linear fit obtained for each choice of the initial phase and phase variation.
The edge charge vanishes in the thermodynamic limit $L\rightarrow\infty$.
}
\label{fig:Scaling}
\end{figure}%

In this appendix we detail the calculation of the variation of the center of charge $\Delta\langle r \rangle$ and the edge charge $\sigma$ in \cref{eq:CCQ} in the thermodynamic limit.
At first, we calculate directly the center of charge $\langle r(\varphi) \rangle=\frac{1}{N}\int_{0}^{L} {\rm d} r \rho(r,\varphi) r$ for different values of the system size $L$. 
For large system sizes, the calculated variation of the center of charge $\Delta\langle r \rangle=\langle r(\varphi) \rangle-\langle r(\varphi_0) \rangle$ is linear in $1/L$. 
To obtain the variation of the center of charge $\Delta\langle r \rangle_\infty$ in the thermodynamic limit, we fit this linear relation as
\begin{equation}
\Delta\langle r \rangle_L = \Delta\langle r \rangle_\infty + \xi/L .
\label{eq:Scaling}
\end{equation}
Therefore we obtain the edge charge in the thermodynamic limit as $\sigma=Q-\nu\Delta\langle r \rangle_\infty$ [see \cref{eq:CCQ}].
The variation of the center of charge and the edge charge for $p/q=1/3$ in the thermodynamic limit (see \cref{fig:Centersigma}) are obtained via \cref{eq:Scaling} with system sizes up to $L=720$. 
In \cref{fig:Scaling} we show the convergence of the edge charge in the thermodynamic limit for $\Delta\varphi=2\pi/3$ and $\Delta\varphi=4\pi/3$.
According to \cref{eq:CCQ,fractionalcharge2}, the edge charge vanishes in the thermodynamic limit for any $\Delta\varphi=n2\pi/q$.

\begin{figure}[t!]
\centering
\includegraphics[scale=1,resolution=600]{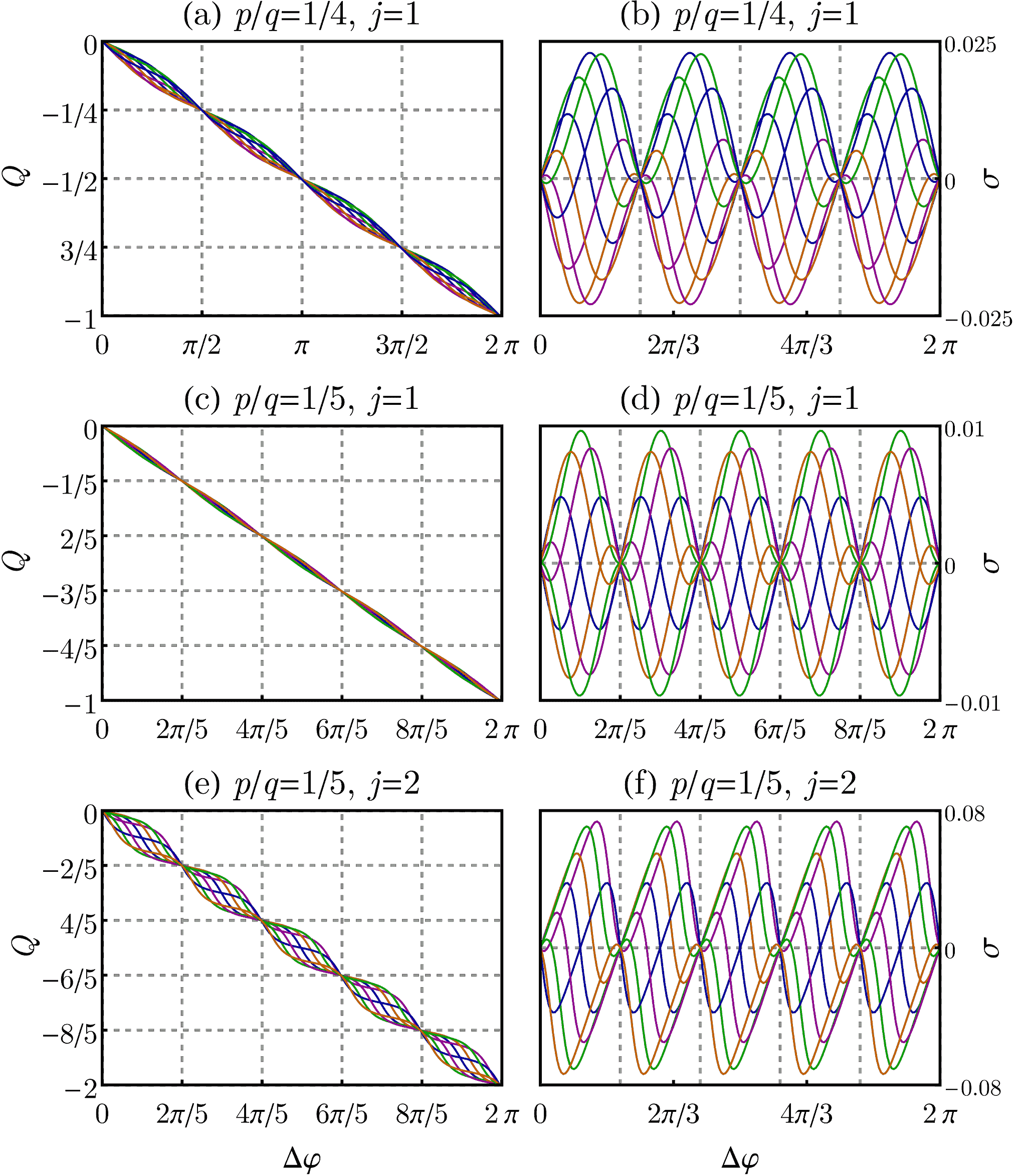}
\caption{
Charge transferred $Q$ and edge charge $\sigma$ in the thermodynamic limit as a function of the modulation phase variation $\Delta\varphi$ for $p/q=1/4$ in the first intraband gap $j=1$ at filling $\nu=1/4$ with Chern number $C_1=-1$ (a-b) and for $p/q=1/5$ in the first $j=1$ (c-d) and second intraband gap $j=2$ (e-f) respectively with filling $\nu=1/5$ and $2/5$, and Chern numbers $C_j=-1$ and $-2$.
The charge transferred is calculated via numerical integration of \cref{ChernPhi}, while the edge charge is obtained from the variation of the center of charge calculated via \cref{eq:Scaling}, with $\Delta\varepsilon/t=2$.
Different curves correspond to different initial phase values $\varphi_0$.
For any $\Delta\varphi=2\pi n/q$, the edge charge vanishes while the charge transferred is quantized as fractions of the Chern number $Q=n C_j/q$, according to \cref{fractionalcharge,fractionalcharge2}.
}
\label{fig:q}
\end{figure}

In \cref{fig:q} we show the pumped charge and the edge charge in the thermodynamic limit in the case $p/q=1/4$ for the first intraband gap and in the case $p/q=1/5$ for the first two gaps.
The pumped charge and the edge charge satisfy the fractional quantization described in \cref{fractionalcharge,fractionalcharge2} for $\Delta\varphi=n2\pi/q$.

\section{Equivalence between the 1D superlattice and the 2D Hofstadter system}
\label{app:hofstadter}

In this appendix we show that the 1D superlattice Hamiltonian~(\ref{TBMomentum}) is equivalent to the Hamiltonian of a 2D Hofstadter Hamiltonian on a rectangular lattice. 
The Hofstadter system is described by a tight-binding model with hopping parameters in the form $t_{ij}=t e^{-\frac\imath\hbar\int{{\bf A}\cdot d{\bf l}}}$, where the integral is along the hopping path~\cite{Peierls1933}.
For a uniform magnetic field with magnitude $B$ perpendicular to the lattice plane, the Landau gauge can be chosen such that the complex phase is zero for sites aligned along the $y$ direction, while one has $t_{ij}=t_x e^{\pm\imath 2\pi \beta x_i}$ along the $x$ direction, with the plus sign for $j>i$ hoppings and $\beta={\Phi}/{\Phi_0}$, where $\Phi$ is the magnetic flux per lattice cell, and $\Phi_0=h/e$ the magnetic flux quantum. 
Assuming that the magnetic field is strong enough for electron spins to be completely aligned, the Hofstadter Hamiltonian in momentum space reads
\begin{equation}
\label{Hofstadter}
{\cal H}=\sum_{\bf k} -2t_x\cos{k_x}\ c_{\bf k}^\dag c_{\bf k}^\nodag
- t_y e^{\imath k_y} c_{{\bf k}+{\bf g}}^\dag c_{\bf k}^\nodag
 + \text{h.c.},
\end{equation}
where ${\bf g}=2\pi \beta\hat{\bm{x}}$, $\hat{\bm{x}}$ being the unit vector along the $x$ direction. 
This Hamiltonian is equivalent to the Hamiltonian~(\ref{TBMomentum}) if one sets $\beta=\alpha$, $t_x=t$, $t_y=-\Delta\varepsilon/2$, $k_x=k$, and $k_y=\varphi$. 

\section{Symmetry properties of the 1D Hamiltonian}
\label{app:symmetry}

We show in this appendix that the Hamiltonian~(\ref{TBMomentum}) is periodic in momentum with period $2\pi/q$. 
Since this Hamiltonian couples states with momenta ${k'-k}=2\pi\alpha$, the time evolution of a Bloch state $e^{-\imath {\cal H}t/\hbar}c^\dag_{k}$ is a linear combination of states with momenta $k+{2\pi n\alpha}$, with $n$ integer. 
If the system is commensurate, i.e., if $\alpha$ is rational, the number of these Bloch states is finite. 
In fact, if $\alpha=p/q$ with $p$ and $q$ coprimes, one has $k+{2\pi q\alpha}=k+{2\pi p}\equiv{k}$, which implies that there are only $q$ states with distinct momenta ${k}+{2\pi n\alpha}$, i.e., those with $0\le n\le q-1$, which form a lattice in the reciprocal space spanned by $2\pi\alpha$. 
As a consequence the Hamiltonian is separable as ${\cal H}=\sum_{k} {\cal H}_{k}$ with
\begin{align}
{\cal H}_k=&\sum_{n} 
-2t\cos{(k+2\pi n\alpha)}\ c_{k+2\pi n\alpha}^\dag c_{k+2\pi n\alpha}^\nodag
\nonumber\\
+&\sum_{n} 
\frac{\Delta\varepsilon}{2}e^{\imath\varphi}c_{{k+2\pi (n+1)\alpha}}^\dag c_{k+2\pi n\alpha}^\nodag+\text{h.c.}.
\label{TBseparated}
\end{align}
By direct substitution in \cref{TBseparated} it can be seen that ${\cal H}_{k}={\cal H}_{k+2\pi\alpha}$, i.e., the Hamiltonian is periodic in momentum with period $2\pi\alpha$. 
However, since the Diophantine equation $n p+m q=1$ has infinite many solutions $n$ and $m$ for any coprimes $p$ and $q$, one has $2\pi n\alpha=2\pi/q- 2\pi m\equiv2\pi/q$. 
Therefore the system is invariant upon the transformation $k\rightarrow k+2\pi/q$, i.e., it is periodic in momentum with period $2\pi/q$. 

The periodicity of the Hamiltonian~(\ref{TBMomentum}) in modulation phase $\varphi$ can be also inferred from the equivalence of the 1D superlattice and the 2D Hofstadter system. 
In the Hofstadter system, the uniform magnetic field perpendicular to the 2D lattice is described by a vector potential whose direction can be chosen, due to the gauge invariance, arbitrarily in the lattice plane. 
This gauge invariance implies that the two components of the momentum are equivalent upon rotations. 
Correspondingly, the Hamiltonian~(\ref{TBMomentum}) of the 1D system is self-dual~\cite{Johansson1990} under the transformation $t\leftrightarrow-\Delta\varepsilon/2$ and $k\leftrightarrow\varphi$. 
Hence, the periodicity of the Hamiltonian in the momentum $k$ implies the periodicity in the modulation phase $\varphi$ with the same period $2\pi/q$. 


\begin{thebibliography}{10}
\bibitem{note_wannier}
The charge density in Fig.~\ref{fig:Density} has been calculated via exact
  diagonalization of the Hamiltonian (\ref{TBSpace}) on the basis of maximally
  localized Wannier functions $\phi_i(r)\propto\sum_k e^{\imath k(r-R_i)}u(r)$,
  with $u(r)$ an arbitrary periodic function with the same period of the
  lattice.
\end{thebibliography}

\end{document}